\newcommand{\be}{\begin{equation}}\newcommand{\ee}{\end{equation}}
\newcommand{\bea}{\begin{eqnarray}}\newcommand{\eea}{\end{eqnarray}}
\newcommand{\p}[1]{(\ref{#1})}
\documentstyle[12pt]{article}
\begin{document}

\thispagestyle{empty} \begin{flushright}    ENSLAPP-L-371/92 \\
 BONN-HE-92-11 \\
 April 1992 \end{flushright}
\bigskip\bigskip\begin{center} {\bf \Large{Twistor-like superstrings
with D = 3, 4, 6 target-superspace and N = (1,0), (2,0), (4,0)
world-sheet supersymmetry}}\end{center}  \vskip 1.0truecm
\centerline{\bf F. Delduc$\, ^{*}$, E. Ivanov$\, ^{**}$ and E.
Sokatchev$\, ^{***\;\dag}$} \bigskip \nopagebreak \begin{abstract} We
construct a manifestly $N=(4,0)$ world-sheet supersymmetric
twistor-like formulation of the $D=6$ Green-Schwarz superstring,
using the principle of double (target-space and world-sheet)
Grassmann analyticity. The superstring action contains two Lagrange
multiplier terms and a Wess-Zumino term. They are written down in the
analytic subspace of the world-sheet harmonic $N=(4,0)$ superspace,
the target manifold being too an analytic subspace of the harmonic
$D=6\;\; N=1$ superspace. The  kappa symmetry of the $D=6$
superstring is identified with a Kac-Moody extension of the
world-sheet $N=(4,0)$ superconformal symmetry. It can be enlarged to
include the whole world-sheet reparametrization group if one
introduces the appropriate gauge Beltrami superfield into the action.
To illustrate the basic features of the new $D=6$ superstring
construction, we first give some details about the simpler (already
known) twistor-like formulations of $D=3, N=(1,0)$ and $D=4, N=(2,0)$
superstrings.

\end{abstract} \bigskip \nopagebreak \begin{flushleft} \rule{2
in}{0.03cm}\\ {\footnotesize \ ${}^{*}$  Laboratoire de Physique
Th\'eorique, Unit\'e associ\'ee au CNRS URA 1436, Ecole Normale
Sup\'erieure, 46 all\'ee d'Italie, 69364 Lyon Cedex 07, France.}
\\ {\footnotesize \ ${}^{**}$ Laboratory of Theoretical Physics,
Joint Institute for Nuclear Research, 141980 Dubna near Moscow,
Russia.}
  \\ {\footnotesize \ ${}^{***}$ Physikalisches Institut,
  Universit\"at Bonn, Nussallee 12, 5300 Bonn 1, Germany.}
\\ {\footnotesize \ ${}^{\dag}$  On leave from the Institute for
Nuclear Research and Nuclear Energy, Sofia, Bulgaria.}
\\ \end{flushleft}

\newpage\setcounter{page}1

\section{Introduction}

Recently, essential progress has been made in understanding the
nature of the mysterious kappa symmetry, which is the basic
ingredient of any self-consistent covariant super $p$-brane theory [1
- 10].

The first decisive step in this direction has been made by Sorokin,
Tkach, Volkov and Zheltukhin (STVZ) \cite{{a1},{a00}} in the case of
superparticle ($p=0$). Replacing the superparticle momentum by a new
twistor variable (a commuting spinor), these authors reformulated the
massless $D=3,\; 4$ superparticle actions in a manifestly world-line
and target-space supersymmetric way. In this new formulation, the
local kappa symmetry inherent to the original form of the action
becomes part of the diffeomorphism group of the world-line $N=1$ and
$N=2$ superspace, respectively.  More precisely, it is identified
with the odd sector of the group of local world-line superconformal
transformations  and so it is realized as an off-shell symmetry of
the action. The latter is naturally written down in terms of the
world-line superfields.

These results have been further extended in [3 - 8]. Delduc and
Sokatchev \cite{a2} have put the $D=3,\;4$ STVZ actions in an
arbitrary curved background and have given a similar description for
the massless $D=6$ superparticle.  In the case $D=4$ the key role of
the combined complex Grassmann analyticity (chirality) in the
world-line and target superspaces has been revealed. A new crucial
feature of the $D=6$ action is that it relies on the concept of
double harmonic analyticity which generalizes the complex analyticity
of the $D=4$ case: The target superspace is the analytic harmonic
$D=6$ superspace \cite{a9} and its coordinates are superfields living
in an analytic subspace of the world-line harmonic $N=4$ superspace.
Ivanov and Kapustnikov \cite{a3} have developed a twistor-type
formulation for the massive $D=2$ superparticle, where the
corresponding kappa symmetry could once again be identified with
local superconformal symmetry in the $N=1$ world-line superspace.
Based on this formulation, they have also found a very simple and
efficient algorithm for constructing higher-order supersymmetric and
kappa invariant corrections to the minimal superparticle action.  In
\cite{a4} this result has been extended to the case of the massive
$D=3$ superparticle. Also, it has been observed [5, 6] that the
twistor-like superfield actions of the massive $D=2$, $D=3$
superparticles follow via a dimensional reduction from those of the
massless $D=3$, $D=4$ superparticles and the harmonic superspace
action for the massive $D=5$ superparticle has been derived in this
way from the $D=6$ action of ref. \cite{a2}. A group-theoretical
analysis of the STVZ construction has been undertaken by Galperin,
Howe and Stelle \cite{a5}.  Howe and Townsend \cite{a6} interpreted
the world-line superfield STVZ actions as those of supersymmetric
Chern-Simons mechanics.

Though being very interesting and suggestive in their own right, the
twistor formulation of the superparticle and the new interpretation
of the relevant kappa symmetry should be considered as preparatory
stages before approaching the cases of super $p$-branes with $p \geq
1$ and, first of all, superstrings ($p=1$) \cite{a10}. The first
steps in generalizing this approach to superstrings have been made in
ref. \cite{{a00},{a7}}.  Berkovits \cite{a7} proposed a twistor-like
action for the $N=1$ Green-Schwarz (GS) superstring which has the
same form for all classically allowed superstring dimensions.
However, this action possesses only one manifest off-shell world-line
supersymmetry, while the kappa symmetry of the GS action is known to
involve 1, 2, 4 and 8  Grassmann parameters for  target spaces of
dimension 3, 4, 6 and 10, respectively.  Thus, the formulation of
\cite{a7} could only fully explain kappa symmetry for three
target-space dimensions.  The complete twistor action for the $D=4$
GS superstring, with manifest $N=(2,0)$ world-sheet supersymmetry,
has been found in \cite{a3} \footnote {An $N=(2,0)$ world-sheet
supersymmetric twistor-like formulation of the $D=10$ superstring,
with two of the eight kappa supersymmetries traded for world-sheet
superconformal ones, has been independently and simultaneously
proposed by Tonin \cite{a8}.}.  Like Berkovits' action, it is written
as a pure Wess-Zumino term in the world-sheet superspace, however the
latter is now identified with the complex chiral subspace of the
world-sheet $N=(2,0)$ superspace.  Kappa symmetry turns out to be
directly related to a restricted class of diffeomorphisms of the
world-sheet superspace, namely to $N=(2,0)$ superconformal
transformations with the parameters having an
 arbitrary dependence on the remaining (inert under supersymmetry)
world-line bosonic coordinate. The guiding principle in constructing
the twistor description of the $D=4$ superstring, just as in the case
of the $D=4$ superparticle \cite{a2}, was that of double Grassmann
analyticity. The coordinates of the target {\it chiral} $D=4\;\; N=1$
superspace were defined as {\it chiral} superfields with respect to
world-sheet $N=(2,0)$ supersymmetry. In \cite{a3} it has been
suggested that this principle could be applied in twistor-like
formulations of other super $p$-branes with kappa invariance.

The main result of the present article (Section 4) is a twistor
formulation of the $D=6\;\;N=1$ superstring based on a combination of
the ideas underlying the twistor descriptions of the $D=6\;\;N=1$
superparticle and the $D=4\;\;N=1$ superstring. It is given in terms
of the coordinates of the analytic harmonic $D=6$ target superspace,
which are in turn world-sheet $N=(4,0)$ harmonic analytic
superfields.  The action consists of a Wess-Zumino term and two
Lagrange multiplier terms. The latter imply on-shell constraints on
the superfields, which restrict their harmonic dependence. We analyse
the component content of the theory and show that it is equivalent to
the GS superstring in six dimensions. The superstring kappa symmetry
is identified with the odd sector of the $N=(4,0)$ superconformal
group (gauged, as in the case of $D=4$ superstring, by the
supersymmtery-inert world-sheet coordinate). We show that by
introducing a gauge $N=(4,0)$ Beltrami supermultiplet, the action can
be made invariant under an extended world-sheet superdiffeomorphism
group including {\it arbitrary} bosonic world-sheet
reparametrizations.

For convenience of the reader in Sections 2, 3 we recall the already
known examples. In Section 2 we review the simplest case of a
twistor-like superstring, the one with a $D=3$ target superspace and
$N=(1,0)$ world-sheet supersymmetry. There one can already see the
r\^ole played by the twistor variables and the relationship between
kappa symmetry and world-sheet supersymmetry.  We study the
symmetries of the action and we show how to promote the right
conformal invariance to full right diffeomorphisms by introducing a
Beltrami gauge field.  In Section 3 we present a modification of the
$D=4\;\; N=(2,0)$ superstring of ref. \cite{a3}, in which the
Wess-Zumino term is given as a full superspace integral and the
on-shell condition on the superfields is included in the action with
a Lagrange multiplier.

\section{D=3 superstring with N=(1,0) world-sheet supersymmetry }

In this section we review the formulation of twistor-like superstring
theories with $N=(1,0)$ world-sheet supersymmetry. Although this can
be done for the target-space dimensions $D=3,\;4,\;6,\;10$ \cite{a7},
we shall restrict ourselves to the case $D=3$. The reason is that
with just one world-sheet supersymmetry one can fully explain the
kappa symmetry of a three-dimensional superstring only.

\subsection{Superspace action}

The action we shall consider has $N=(1,0)$ supersymmetry, so we
introduce  a Grassmann coordinate $\eta$ besides the light-cone
world-sheet coordinates $\xi^{(+)}, \xi^{(-)}$. The heterotic nature
of $N=(1,0)$ supersymmetry means that we supersymmetrize only one of
the world-sheet directions, e.g., $\xi^{(-)}$. The corresponding
covariant spinor derivative on the world sheet is $D=\partial_\eta
+i\eta\partial_{(-)}$, $D^2=i\partial_{(-)}$. The target superspace
coordinates $X^\mu (x,\eta)$, $\Theta^\alpha (x,\eta)$ are defined as
world-sheet superfields.  $\Theta^\alpha$ is a $D=3$ Majorana spinor.
In the following, heavy use will be made of the
relation\footnote{This identity is  valid in $D=3,\;4,\;6,\;10$. It
is crucial for the consistency of the Green-Schwarz superstring
theories. At the same time, it is in the basis of the twistor-like
approach.} \begin{equation} (\gamma^\mu)_{\alpha
(\beta}(\gamma_\mu)_{\delta\epsilon)} =0\;.  \label{relg}
\end{equation} The action describing the $D=3$ superstring is
\begin{equation} S=\int d^2\xi d\eta\left[ \Pi_{(+)\mu}
D\Theta\gamma^\mu\Theta -iP_\mu (DX^\mu-iD\Theta\gamma^\mu\Theta)
\right],\label{action0}\end{equation}
 where \begin{equation} \Pi_{(+)}^\mu=\partial_{(+)}X^\mu
-i\partial_{(+)}\Theta\gamma^\mu\Theta \; , \end{equation} and
$P_\mu$ is a Lagrange multiplier superfield. This action is invariant
under global space-time supersymmetry transformations,
$$\delta\Theta^\alpha=\epsilon^\alpha\;,\ \ \delta
X^\mu=i\Theta\gamma^\mu\epsilon \;. $$ Indeed, up to total
derivatives and making use of the identity (\ref{relg}), one can
write down the variation of the first term $S_{1}$ of this action as
follows $$\delta S_{1} =\int d^2\xi d\eta \;
(DX^\mu-iD\Theta\gamma^\mu\Theta)\;\partial_{(+)}
\Theta\gamma_\mu\epsilon\; . $$ Clearly, this can be absorbed into a
variation of the Lagrange multiplier $P_\mu$ (the constraint
introduced by $P_\mu$ is invariant in its own right).

The action (\ref{action0}) is also invariant under a restricted class
of left superdiffeomorphisms:  \be \delta\xi^{(-)} = \Lambda^{(-)} -
{1\over 2}\eta D\Lambda^{(-)}, \ \ \ \delta\eta = -{i\over 2}
D\Lambda^{(-)}.  \label{left1}\ee (the world-sheet coordinate
$\xi^{(+)}$ is not affected by \p{left1}). They  do not change the
form of the spinor covariant derivative: $$ \delta D = -{1\over 2}
\partial_{(-)}\Lambda^{(-)}\; D\;.  $$ This transformation law is
reminiscent of a superconformal transformation. However, the
$\xi^{(+)}$-dependence of the superfield parameter
$\Lambda^{(-)}(\xi^{(+)},\xi^{(-)},\eta)$ is not restricted and so
these transformations actually constitute a Kac-Moody extension of
the $N=(1,0)$ superconformal group. Their active form on any scalar
superfield $\phi$ is \begin{equation} \delta\phi
=\Lambda^{(-)}\partial_{(-)}\phi - {i\over 2}D\Lambda^{(-)}D\phi\; .
\label{ldiff0} \end{equation}

Taking for the time being all fields to transform as scalars, one
finds, up to total derivatives and using again eq. (\ref{relg}):
$$\delta S =-i\int d^2\xi d\eta\;
(DX^\mu-iD\Theta\gamma^\mu\Theta)\left[ D \;(
\partial_{(+)}\Lambda^{(-)}\; D\Theta\gamma_\mu\Theta) -{1\over
2}\;\partial_{(+)}D\Lambda^{(-)}\; D\Theta\gamma_\mu\Theta
\right]\;.  $$ It is thus clear that, like in the previous case, one
can choose an additional variation of the Lagrange multiplier so that
the action (\ref{action0}) will be invariant.

Finally, the action (\ref{action0}) is invariant under right
conformal transformations:  $$ \delta\xi^{(+)} =
\Lambda^{(+)}(\xi^{(+)})\;, $$ the superfields $X^\mu$ and $\Theta$
being scalars, and the Lagrange multiplier $P_\mu$ a density.  The
fact that this is not a full diffeomorphism invariance is an
indication that one of the two Virasoro constraints does not follow
from the action. This conformal invariance can be promoted to a full
right diffeomorphism invariance, $\Lambda^{(+)} =
\Lambda^{(+)}(\xi^{(+)},\xi^{(-)},\eta)$ by introducing a new field
which will at the same time generate the missing constraint. To
clarify this procedure, one could consider a simplified example. Take
a bosonic string with the following action:  \be S=\int d^2\xi
\left(\partial_{(+)}X^{\mu}\partial_{(-)}X_{\mu}
+\nu\partial_{(-)}X^{\mu}\partial_{(-)}X_{\mu}\right).  \label{mu}\ee
This action has the same type of invariances as (\ref{action0}),
namely
 full left diffeomorphisms and right conformal invariance. Note that
 the Lagrange multiplier $\nu$ is the gauge field for left
 diffeomorphisms. At the same time, it generates the Virasoro
 constraint $\partial_{(-)}X^{\mu}\partial_{(-)}X_{\mu}=0$. To
restore the right diffeomorphisms, one makes  an arbitrary change of
coordinate from $\tilde\xi^{(+)}$ to a new coordinate
$\xi^{(+)}=\xi^{(+)}(\tilde\xi^{(+)},\xi^{(-)})$. The Jacobian of
this transformation may be reabsorbed into a rescaling of the
Lagrange multiplier $\nu$. The only change is thus in the derivative
$$\tilde\partial_{(-)}={\cal D}_{(-)}=\partial_{(-)}+
\mu(\xi)\partial_{(+)}, \ \ \ \mu(\xi)=
\tilde\partial_{(-)}\xi^{(+)}.$$ The new field $\mu$ is the gauge
field for the right diffeomorphisms.  Further, the variation with
respect to $\mu$ of the action (\ref{mu}), with $\partial_{(-)}$
replaced by ${\cal D}_{(-)}$,  produces the second Virasoro
constraint \be \label{v2}
\partial_{(+)}X^{\mu}\partial_{(+)}X_{\mu}+2\nu\partial_{(+)}X^{\mu}
{\cal D}_{(-)}X_{\mu}=0.\ee Effectively, the fields $\mu$ and $\nu$
are two of the components of the world-sheet metric. The third one is
not present because it corresponds to a Weyl rescaling.  In an
analogous way, we can covariantize the derivatives $D$,
$\partial_{(-)}$ in the superstring action (\ref{action0})  as
follows:  \begin{eqnarray} D\rightarrow {\cal D}=D+\chi\partial_{(+)}
&&\nonumber\\ \partial_{(-)}\rightarrow {\cal D}_{(-)}= -i{\cal D}^2
\;,&&\end{eqnarray} where the superfield $\chi$ transforms under
right diffeomorphisms:  \begin{equation}
\delta_{right}\chi=-\Lambda^{(+)}\partial_{(+)} \chi+{\cal
D}\Lambda^{(+)}\ \ \  \Lambda^{(+)} =
\Lambda^{(+)}(\xi^{(+)},\xi^{(-)},\eta)\;.  \label{rdiff}
\end{equation}

The action (\ref{action0}) with these replacements made is still
invariant under global space-time supersymmetry, as well as under
left superdiffeomorphisms (\ref{left1}) provided the derivatives in
the transformation law (\ref{ldiff0}) are replaced by covariant
derivatives, the transformation law for the Lagrange mutiplier
$P_\mu$ is suitably modified and $\chi$ is inert under (\ref{left1})
 \footnote{These transformation laws of $\chi$ are compatible with
the Lie bracket structure :  $ \left[ \delta_{left},\;\delta_{right}
\right] \sim \delta_{right}\;.$} \be\label{csc3} \delta_{left} \chi =
0\;.  \ee

Note that the so covariantized left superdiffeomorphisms, being
rewritten in a passive form, involve an induced field-dependent
transformation of the coordinate $\xi^{(+)}$. In components this
leads to a non-standard form of the world-sheet reparametrizations.
Besides, these transformations close with a field-dependent bracket
parameter.  By making use of the invariance under $\xi^{(+)}$
diffeomorphisms one may ``subtract'' this unwanted induced shift of
$\xi^{(+)}$ from the left diffeomorphisms and restore the original
Lie bracket structure.  As a result of this redefinition, the
variation $\delta_{left}\chi$ becomes \begin{equation}
\tilde{\delta}_{left}\chi=\tilde{\Lambda}^{(+)}\partial_{(+)}
\chi-{\cal D} \tilde{\Lambda}^{(+)}\;, \;\;\;\;\;\;\;
\tilde{\Lambda}^{(+)} = i ( \Lambda^{(-)}{\cal D}\chi + \frac{1}{2}
{\cal D} \Lambda^{(-)} \chi )\;.  \label{rconf} \end{equation} We
point out that $\delta_{left}$ and $\tilde{\delta}_{left}$ coincide
modulo a $\xi^{(+)}$ diffeomorphism transformation.

In fact, the superfield $\chi$ is a pure gauge. In particular, using
the freedom in the parameter $\Lambda^{(+)}$, one can fix a
Wess-Zumino gauge (and we shall assume it in what follows), where the
only surviving component of $\chi$ is the Beltrami parameter
$\mu(\xi)=-iD\chi\vert_{\eta=0}$. Note that this does not restrict
the bosonic part of the world-sheet reparametrization.

We would like to point out that there is an alternative approach, in
which one could have started with the complete formalism of $N=(2,0)$
supergravity on the world sheet (see, for example, \cite{a8}). There
one introduces the full set of zweibeins and connections. However,
most of them simply drop out from the twistor-like action
\p{action0}. As we have just seen, the only gauge superfield really
needed for maintainig the gauge symmetries (local supersymmetry and
diffeomorphisms) is the Beltrami superfield.

\subsection{Component action. World-sheet conformal supersymmetry
versus kappa symmetry}

 We shall denote the physical components of the superfields $X$,
$\Theta$ and $P$ by: $$x^\mu(\xi)=X^\mu\vert_{\eta=0}\;,\ \
\theta^\alpha(\xi)=\Theta^\alpha\vert_{\eta=0}\;,\ \
\lambda(\xi)=D\Theta^\alpha\vert_{\eta=0}\;,\ \
p_\mu(\xi)=P_\mu\vert_{\eta=0}$$  (we discard some purely auxiliary
fields which are expressed in terms of the physical ones on shell).
Further, we introduce the notation:
$$\pi^\mu_{(+)}=\partial_{(+)}x^\mu-i
\partial_{(+)}\theta\gamma^\mu\theta ,$$ $$\hat{\pi}^\mu_{(-)}={\cal
D}_{(-)}x^\mu-i {\cal D}_{(-)}\theta\gamma^\mu\theta \equiv
\pi^{\mu}_{(-)} + \mu \pi^\mu_{(+)}\; ,$$ with ${\cal
D}_{(-)}=\partial_{(-)}+ \mu\partial_{(+)}$.  Then it is not hard to
obtain the component form of the action (\ref{action0}):  \be S=\int
d^2\xi\;\left(\pi_{(+)\mu }\;\lambda\gamma^\mu\lambda +i\;\pi_{(+)\mu
}\; {\cal D}_{(-)}\theta\gamma^\mu\theta
-i\;\lambda\gamma_\mu\lambda\;\partial_{(+)}\theta \gamma^\mu\theta
 +p_\mu\; (\hat{\pi}^\mu_{(-)}-\lambda\gamma^\mu\lambda) \right)\;.
\label{local}\ee The left superdiffeomorphisms \p{ldiff0}
(covariantized with respect to $\xi^{(+)}$ diffeomorphisms) contain,
in particular, local left supersymmetry with parameter $\rho=-{i\over
2}D\Lambda^{(-)}\vert_{\eta=0}$. It acts on the above fields  in the
following way: $$ \delta x^\mu =i\rho\;\lambda\gamma^\mu\theta\;,\ \
\delta \theta^\alpha =\rho\;\lambda^\alpha \;, $$ \be
\delta\lambda^\alpha =i\rho\;{\cal D}_{(-)}\theta\;,\ \ \delta p_\mu
=i\partial_{(+)}\left( \rho\;\lambda\gamma_\mu\theta \right)\;.
\label{loccon}\ee

The field $p_\mu$ is a Lagrange multiplier for the constraint
\begin{equation} \hat{\pi}^\mu_{(-)} =\lambda\gamma^\mu\lambda\;.
\label{stv0} \end{equation} With the help of the identity \p{relg}
this implies that $\hat{\pi}^\mu_{(-)}$ is a light-like vector:
\begin{equation} \hat{\pi}^\mu_{(-)}\hat{\pi}_{\mu(-)}=0\;.
\label{light0} \end{equation} In fact, eq. \p{light0} is one of the
two Virasoro constraints for the superstring (the second Virasoro
constraint can be obtained by varying the Beltrami parameter $\mu$ in
\p{local}, see below). Here we see the main idea of the twistor
approach in action: A light-like vector is represented as a pair of
commuting spinors (twistor variables).  Further, we note that the
twistor variable $\lambda^\alpha$ appears in \p{local} only in the
combination $\lambda\gamma^\mu\lambda$, so (\ref{stv0}) may be used
to eliminate it from the action, which then becomes:
\begin{equation} S=\int d^2\xi\;\left(\pi_{\mu
(+)}\hat{\pi}^\mu_{(-)} +i\;\pi_{\mu (+)}\;
\partial_{(-)}\theta\gamma^\mu\theta -i\;\hat\pi_{\mu (-)}\;
\partial_{(+)}\theta \gamma^\mu\theta\right)\;.
\label{GS}\end{equation} This action, accompanied by the constraint
(\ref{light0}), is just the action of the GS superstring \cite{a10}.
As a consequence of the elimination of the  twistor-like variable
$\lambda^\alpha$, the action \p{GS} has lost the local left
supersymmetry \p{loccon} of the action \p{local}.  Instead, it has a
new local symmetry, \be\label{kappa} \delta x^\mu =
i\kappa\gamma^\nu\gamma^\mu\theta\; \hat{\pi}_{\nu(-)}\;, \ \ \
\delta\theta^\alpha = (\kappa\gamma^\mu)^\alpha \hat{\pi}_{\mu(-)}\;,
\ee which is just the familiar kappa symmetry of the GS superstring.
Actually, the transformations \p{kappa} can be recast in the form of
local supersymmetry \p{loccon}, if one defines $\rho =
\lambda^\alpha\kappa_\alpha$, and then uses the {\it on-shell
condition} \p{stv0}, as well as the Fierz identity for the $D=3$
gamma matrices. This procedure shows that kappa symmetry is
equivalent to world-sheet supersymmetry only on shell (and hence the
on-shell closure of the algebra of kappa symmetry). Further, we
recall the well-known fact  that because of the presence of the
light-like vector $\hat{\pi}_{(-)}$ in \p{kappa} only half of
$\kappa^\alpha$ are true gauge parameters (they are used to  gauge
away half of $\Theta^\alpha$). In the allowed dimensions
$D=3,\;4,\;6,\;10$ that half of $\kappa^{\alpha}$ can be matched by
$N=(1,0),\;(2,0),\;(4,0),\;(8,0)$ world-sheet supersymmetries.  In
this section we consider $N=(1,0)$, so it was natural to associate it
with the $D=3$ superstring. As a matter of fact, the whole discussion
above applies to any of the dimensions $D=3,\;4,\;6,\;10$, except for
the relationship between local world-sheet supersymmetry and kappa
symmetry.

Let us explain in more detail how the second Virasoro constaint
follows from the action \p{local}.  After a redefinition of the
Lagrange multiplier as $$ p^{\mu} = \tilde{p}^{\mu} -
i\;\partial_{(+)}\theta\gamma^{\mu}\theta + \pi_{(+)}^{\mu}\;, $$ the
action takes the form \be\label{local1} S=\int
d^2\xi\;\left(\pi_{(+)\mu}\hat{\pi}_{(-)}^{\mu} +i\;\pi_{(+)\mu}\;
{\cal D}_{(-)}\theta\gamma^\mu \theta
-i\;\hat\pi_{(-)\mu}\;\partial_{(+)}\theta\gamma^\mu \theta
 +\tilde{p}_\mu\; (\hat{\pi}^\mu_{(-)}-\lambda\gamma^\mu\lambda)
\right)\;.  \ee Varying \p{local1} with respect to $\tilde{p}_{\mu}$
leads to the already known twistor constraint \p{stv0}, whose
corollary is eq. \p{light0}, while varying with respect to $\mu$
gives \be\label{rvc1} (\tilde{p}_{\mu} +
\pi_{(+)\mu})\;\pi_{(+)}^{\mu} = 0 \;.  \ee Let us now look at the
equation of motion for $\lambda^\alpha$ $$ \tilde{p}_{\mu}
(\gamma^{\mu} \lambda)_\alpha = 0\;.  $$ It is well known [1, 2, 6,
7] that for $D=3$ (as well as for $D=4,\;6,\;10$) this equation has
the general solution \be\label{tws} \tilde{p}^{\mu} = c
\lambda\gamma^{\mu}\lambda\;, \ee where $c$ is an arbitrary scalar
field on the world sheet.  Comparing \p{tws} with \p{stv0} we find $$
\tilde{p}^{\mu} = c \hat{\pi}_{(-)}^{\mu}\;.$$ Further, substituting
this into into eq.\p{rvc1} we see that \p{rvc1} is just the second
Virasoro constraint \p{v2}, provided one identifies $$ c = 2\nu\;. $$

We conclude this section by the remark that the first term in the
action (\ref{action0}) is in fact a typical Wess-Zumino
term\footnote{The presence of a WZ term is a characteristic feature
of the GS superstring \cite{mez}.}. This becomes clear after
rewriting the action in the form:  \begin{equation} S=\int d^2\xi
d\eta\left[ ({\cal D}\Theta^\alpha\partial_{(+)}X^\mu -{\cal
D}X^\mu\partial_{(+)}\Theta^\alpha) (\gamma_\mu\Theta)_\alpha -iP_\mu
({\cal D}X^\mu-i{\cal D}\Theta\gamma^\mu\Theta) \right]\;,
\end{equation}
 obtained  by a redefinition of the Lagrange multiplier $P_\mu$.
Indeed, the first term now looks like a Wess-Zumino term $\int
\partial Z^M \partial Z^N B_{NM}(Z)$, where the only non-vanishing
component of the two-form is $B_{\mu\alpha} =
(\gamma_\mu\Theta)_\alpha$.

\section{D=4 superstring with N=(2,0) world-sheet supersymmetry}

The formulation of the $D=4\;\; N=(2,0)$ superstring presented in
this section is equivalent to the one of ref. \cite{a3}, the main
difference is that the WZ term is given as a full world-sheet
superspace integral, and not as a chiral one, as in \cite{a3}. This
allows us to introduce  the on-shell constraints on the superfields
$X$ and $\Theta$ in the action via a Lagrange multiplier, so the new
action involves only unconstrained objects. At the end of this
section we show how the formulation of \cite{a3} can be obtained from
the present one.

We begin by introducing an $N=(2,0)$ world-sheet superspace with
coordinates $\xi^{(+)}, \xi^{(-)}, \eta, \bar\eta$. As before,
supersymmetry acts only in the direction of the coordinate
$\xi^{(-)}$. Further, following the principle of double Grassmann
analyticity formulated in the Introduction, we  introduce the
coordinates of the left and right chiral bases in $D=4$ superspace,
$X^\mu_L$, $\Theta^\alpha$ and $X^\mu_R = \overline{X^\mu_L}$,
$\bar\Theta^{\dot\alpha} = \overline{\Theta^\alpha}$,  as left and
right-handed chiral world-sheet superfields, respectively:
\be\label{chirality} \bar D X^\mu_L = \bar D \Theta^\alpha = 0\;,
\ \ \ D X^\mu_R = D \bar\Theta^{\dot\alpha} = 0\;, \ee where $$ D =
\partial/\partial\eta + i\bar\eta\partial/\partial\xi^{(-)}\;, \ \
\bar D = -\partial/\partial\bar\eta -
i\eta\partial/\partial\xi^{(-)}\;.  $$ The usual, real coordinate
$X^\mu$ is identified with the real part of the chiral ones, $$ X^\mu
= {1\over 2}(X^\mu_L + X^\mu_R), $$ after restricting the imaginary
part to be \be\label{onshell} {i\over 2}(X^\mu_L - X^\mu_R) =
-\Theta\sigma^\mu\bar\Theta \;.  \ee The constraint \p{onshell} is in
fact an equation of motion (see \cite{a2} for the analogous case of
the $D=4\;\; N=2$ superparticle), so we are going to introduce it in
the action with a Lagrange multiplier (cf. \p{action0}).

We propose the following action for the $D=4\;\; N=(2,0)$
superstring:  $$ S ={1\over 2} \int d^2\xi d\eta d\bar\eta
\Big[(\partial_{(+)} X^\mu_L + \partial_{(+)} X^\mu_R
-i\partial_{(+)}\Theta\sigma^\mu\bar\Theta +
i\Theta\sigma^\mu\partial_{(+)}\bar\Theta) \Theta\sigma_\mu\bar\Theta
$$ \be + P_\mu({i\over 2}(X^\mu_L - X^\mu_R) +
\Theta\sigma_\mu\bar\Theta) \Big] \;. \label{action4} \ee Comparing
this action  with that from \cite{a3} (see \p{chiralWZ}), we see that
the main difference is in the first (WZ) term in \p{action4}. Note
also that the Lagrange multiplier term in \p{action4} has exactly the
same form as the action for the $D=4\;\; N=2$ superparticle of ref.
\cite{a2}.

The action \p{action4} has several symmetries.

Firstly, it has global target-space supersymmetry. In the left-handed
chiral basis of $D=4$ superspace it is realized as follows:
\be\label{susy4} \delta X^\mu_L = 2i\Theta\sigma^\mu\bar\epsilon\;,
\ \ \ \delta\Theta^\alpha = \epsilon^\alpha\;, \ee and similarly for
the right-handed basis. The invariance of the Lagrange multiplier
term in \p{action4} is obvious (for the moment, we do not vary
$P^{\mu}$). To check the invariance of the WZ term one has to use the
chirality conditions \p{chirality}, the Fierz identity for the
matrices $\sigma^\mu$ and to ascribe the following transformation law
to the Lagrange multiplier: $$ \delta P^\mu = 2i (\partial_{(+)}
\Theta\sigma^\mu\bar\epsilon - \epsilon \sigma^\mu \partial_{(+)}
\bar\Theta) \;.  $$

Secondly, the action \p{action4} is invariant under restricted left
superdiffeomorphisms:  \be\label{left4} \delta\xi^{(-)}
=\Lambda^{(-)} +{1\over 2} \bar\eta\bar D \Lambda^{(-)} + {1\over
2}D\Lambda^{(-)}\eta\;, \ \  \delta\eta = {i\over 2} \bar D
\Lambda^{(-)}\;, \ \  \delta\bar\eta = -{i\over 2} D \Lambda^{(-)}\;,
\ee where $\Lambda^{(-)}\;(\xi^{(+)},\xi^{(-)},\eta,\bar\eta) =
\overline{\Lambda^{(-)}}$.  These transformations leave the volume of
the real world-sheet superspace invariant, $$ \delta (d^2\xi d\eta
d\bar\eta) = 0\;, $$ and transform the left and right-handed
coordinates of the target superspace as scalars. Consequently, one
finds $$ \delta (\partial_{(+)} X^\mu_L) = -{i\over 2} (\bar D
\partial_{(+)} \Lambda) DX^\mu_L - (\partial_{(+)}\Lambda)
\partial_{(-)}X^\mu_L\;, $$ etc.  Using all this, as well as the
on-shell relation \p{onshell} (in other words, this means finding
appropriate compensating transformations of the Lagrange multiplier),
one can show that the WZ term in \p{action4} is invariant up to total
derivatives. The Lagrange multiplier term is manifestly invariant
too. Like in the case $D=3$, the transformations \p{left4} constitute
an $N=(2,0)$ world-sheet superconformal group with the parameters
local the supersymmetry-inert coordinate $\xi^{(+)}$. Their relation
to kappa supersymmetry is precisely the same as in the case of $D=3$
superparticle (details can be found in \cite{a3}). Note that
\p{left4} leave invariant the chiral subspace $(\xi^{(-)}_{L} =
\xi^{(-)} - i\bar\eta \eta\;, \eta)$ and can be regarded as a
particular class of general diffeomorphisms of the latter,
\be\label{left5} \delta \xi^{(-)}_{L} = \tilde {\Lambda}^{(-)}
(\xi^{(-)}_{L}, \eta, \xi^{(+)}) \;, \;\;\;\;\;\delta \eta = \omega
(\xi^{(-)}_{L}, \eta, \xi^{(+)})\;, \ee which preserve the flat
definition of the world-sheet chiral subspace \be\label{chir} Im
\;\xi^{(-)}_{L} = -i\bar{\eta}\eta\;.  \ee

Finally, the action \p{action4} is obviously invariant under right
conformal reparametrizations of $\xi^{(+)}$. In the end of this
section we shall sketch how these can be promoted to general ones.

Here we shall not investigate the component content of the action
\p{action4}. Instead, we shall show how it can be reduced to the
chiral action of ref. \cite{a3}, where the issue of components has
been discussed. Firstly, we impose the equation of motion
\p{onshell}. Thus the Lagrange multiplier term in \p{action4} drops
out, but the fields become constrained. Secondly, we convert one of
the Grassmann integrations, e.g., $d\bar\eta$ into a spinor
derivative, $\bar D$. Using the relation $\bar D X^\mu_R = 2i
\Theta\sigma^\mu\bar D \bar \Theta$ following from \p{onshell} and
integrating $\partial_{(+)}$ by parts, we obtain the chiral form of
the WZ term from \cite{a3}:  \be\label{chiralWZ} S_{WZ} = - \int
d\xi^{(+)}d\xi^{(-)}_L d\eta (\partial_{(+)} X^\mu +
i\Theta\sigma^\mu \partial_{(+)}\bar \Theta - i \partial_{(+)}
\Theta\sigma^\mu \bar\Theta) \Theta\sigma_\mu\bar D\bar\Theta \;.
\ee Note that the chirality of the integrand in \p{chiralWZ} and the
reality of the action are not manifest, but they follow from the
on-shell relation \p{onshell}.

The  WZ term \p{chiralWZ} has the standard geometric form $\int
\partial Z^M \partial Z^N B_{NM}$, where the two-form is represented
by $B_{\mu\dot\alpha} = (\Theta\sigma_\mu)_{\dot\alpha}$. In
contrast, what we find in the real form \p{action4} of the WZ term is
not the two-form itself, but its chiral-basis prepotential
$\Theta\sigma_\mu \bar\Theta$,  $B_{\mu\dot\alpha} =-\bar
D_{\dot\alpha} (\Theta\sigma_\mu\bar\Theta)$.

Finally, we outline very briefly how the $\xi^{(+)}$ conformal
invariance of the $D=4$ action \p{action4} can be extended to  full
diffeomorphisms.  Like in the case $D=3$, this can be done by
introducing into the action \p{action4} an $N=(2,0)$ Beltrami
superfield. A new point compared to $D=3$ is that the right
superdiffeomorphism group should preserve the notion of world-sheet
chirality. The natural way to satisfy this requirement is to apply
the approach of ref. \cite{a11} to $N=1\;\; D=4$ supergravity in
superspace (for applications to $N=2$ world-sheet geometry in the
context of $N=2$ chiral bosons see also ref. \cite{a12}).  One
changes the coordinate $\xi^{(+)}$ to a complex one $\xi^{(+)}_{L} =
\xi^{(+)} + i \chi^{(+)}(\xi^{(-)}, \xi^{(+)}, \bar{\eta}, \eta)$ and
replaces the conformal shifts of $\xi^{(+)}$ by {\it chiral}
diffeomorphisms of $\xi^{(+)}_{L}$, $$ \delta \xi^{(+)}_{L}  =
\Lambda^{(+)} (\xi^{(+)}_{L}, \xi^{(-)}_{L}, \eta)\;.  $$ Here
$\Lambda^{(+)}$ is an arbitrary complex function of its arguments
and, as before, $\xi^{(-)}_{L} = \xi^{(-)} -i \bar{\eta}\eta$. The
real superfield $\chi^{(+)}$ accomodates the $N=(2,0)$ Beltrami gauge
multiplet. In the Wess-Zumino gauge it reduces to $\chi^{(+)} =
-\bar{\eta}\eta\; \mu (\xi)$ where $\mu$ is the Beltrami parameter,
the same as in the case $D=3$. The covariantized form of the left
diffeomorphisms can be obtained by replacing $\xi^{(+)}$ by
$\xi^{(+)}_{L}$ in the transformation laws \p{left5} and requiring
that the modified transformations still preserve the flat relation
\p{chir}. The resulting transformation laws are nonlinear and
nonpolynomial in $\chi^{(+)}$ but they radically simplify in the
Wess-Zumino gauge.  The only place in the action \p{action4} where
the new Beltrami superfield $\chi^{(+)}$ appears is in the
coordinates of the chiral superfields $X_L$, $X_R$, $\Theta$ and
$\bar\Theta$.  The details of the construction are out of the scope
of our presentation here. We only note that the action considered in
\cite{a3} can be viewed as a particular gauge-fixed form of the
Beltrami-covariantized action, with the whole of $\chi^{(+)}$ gauged
to zero.

\section{D=6 superstring with N=(4,0)  world-sheet supersymmetry}

\subsection{Superspace action and symmetries}

The world-sheet super-coordinates of the $N=(4,0)$ superstring will
be denoted by $\xi^{(+)}$,  $\xi^{(-)}$, $\eta^i$, $\bar\eta_i$ ($i$
is an $SU(2)$ doublet index). To those we add the harmonic
coordinates\footnote{The Lorentz weights $(\pm)$ should not be
confused with the $U(1)$ charges $\pm$.} $u^\pm_i$ of the sphere
$S^2$ (see \cite{a9}). They are used to project the supersymmetric
covariant derivatives: $$\{D_i\;,\bar
D^j\}=i\delta^j_i\partial_{(-)}\ \ \rightarrow \ \ D^\pm = u^\pm_i
D^i\;,\ \ \bar D^\pm = u^\pm_i \bar D^i\;.
 $$ The usual target superspace of the $D=6$ superstring has
coordinates $X^{\alpha\beta} = -X^{\beta\alpha}\equiv
(\gamma_\mu)^{\alpha\beta} X^\mu$ and $\Theta^{i\alpha}$. Here
$\alpha, \beta$ are $SU(4)^*$ spinor indices and $\Theta$ satisfies
the pseudo-Majorana condition $\overline{\Theta^{i\alpha}} =
\epsilon_{ij}C_{\alpha}^{\dot\alpha} \Theta^{i\alpha}$.

As we saw in the preceding section, the $D=4$ superstring is
formulated in terms of the coordinates  of the chiral subspaces of
the target superspace. At the same time, they are taken as chiral
superfields with respect to world-sheet supersymmetry.  This is
actually the principle of double Grassmann analyticity mentioned in
the Introduction. In the case $D=6$ the notion corresponding to $D=4$
chirality is that of $SU(2)$ harmonic Grassmann analyticity
\cite{a9}. Following this idea,\footnote{The same approach proved
successful in the case of the $D=6$ superparticle \cite{a2}.} we
choose to formulate the $D=6$ superstring in terms of the coordinates
$X^{\alpha\beta}(\xi,\eta^+,u)$, $\Theta^{+\alpha}(\xi,\eta^+,u)$
(where $\widetilde{\Theta^{+\alpha}}=C_{\alpha}^{\dot\alpha}
\Theta^{+\alpha}$), defined as analytic harmonic superfields: $$D^+
X^{\alpha\beta}=\bar D^+ X^{\alpha\beta}=0\;,\ \ D^+
\Theta^{+\alpha}=\bar D^+ \Theta^{+\alpha}=0\; .$$ Such superfields
can be considered as unconstrained objects in the analytic subspace
$\xi^{(+)}\;,$ $\xi^{(-)}_A = \xi^{(-)} + i\eta^i\bar\eta^j
u^+_{(i}u^-_{j)}\;,$ $\eta^+ = u^+_i\eta^i\;,$ $\bar\eta^+ =
u^+_i\bar \eta^i$.  Later on, after imposing  the harmonic equations
of motion, the usual coordinate $\theta^{i\alpha}$ will reappear as
the first term of the harmonic expansion of $\Theta^{+\alpha}$.

We shall consider the following action for the $D=6\;\; N=(4,0)$
superstring:  $$ S=\int d^2\xi [du]d^2\eta^+\left[ \partial_{(+)}
X^{\alpha\beta} \Theta^{+\gamma}\Theta^{+\delta}
\epsilon_{\alpha\beta\gamma\delta}\right.  $$ \be \left.
+P_{\alpha\beta}(D^{++}X^{\alpha\beta}-i
\Theta^{+\alpha}\Theta^{+\beta}) +Q^-_\alpha
D^{++}\Theta^{+\alpha}\right]\;.  \label{action} \ee Here $D^{++} =
u^{+i}\partial/\partial u^{-i} + i\eta^+\bar\eta^+\partial_{(-)}$ is
the analytic basis form of the harmonic covariant derivative. The
superfields $P_{\alpha\beta}$ and $Q^-_\alpha$ are analytic Lagrange
multipliers restricting the $u$-dependence of the on-shell fields.
The last two terms in \p{action} are exactly the same as in the case
of the $D=6\;\; N=4$ superparticle (see \cite{a2}).

This action has the three symmetries we already encountered in the
preceding Sections. These are:

1) Global space-time supersymmetry: $$\delta
X^{\alpha\beta}=i(\epsilon^{-\alpha}
\Theta^{+\beta}-\epsilon^{-\beta}
\Theta^{+\alpha})\;,\ \ \delta\Theta^{+\alpha}= \epsilon^{+\alpha}$$
with $\epsilon^{\pm\alpha}=u^\pm_i\epsilon^{i\alpha}$, and
$\epsilon^{i\alpha}$ is a $SU(2)$-Majorana spinor parameter.  Up to
total derivatives, the variation of the action (\ref{action}) is (for
the moment, $P_{\alpha\beta}$ and $Q^{-}_{\alpha}$ are not varied):
$$ \delta S = \int d^2\xi [du]d^2\eta^+\left\{
\left[2(D^{++}X^{\alpha\beta}-i
\Theta^{+\alpha}\Theta^{+\beta})\epsilon^{-\gamma}
\partial_{(+)}\Theta^{+\delta}\right.\right.  $$ \be \left.\left.
-2\partial_{(+)} X^{\alpha\beta}\epsilon^{-\gamma}
D^{++}\Theta^{+\delta}\right] \epsilon_{\alpha\beta\gamma\delta}
+2iP_{\alpha\beta}\epsilon^{-\alpha} D^{++}\Theta^{+\beta}\right\}\;,
\ee which may be compensated by the following variations of the
Lagrange multipliers: $$\delta
P_{\alpha\beta}=-2\epsilon_{\alpha\beta\gamma\delta}\;
\epsilon^{-\gamma} \partial_{(+)}\Theta^{+\delta}\;,\ \ \delta
Q^-_\alpha =-2 \epsilon_{\alpha\beta\gamma\delta}\;
\partial_{(+)}X^{\beta\gamma}\epsilon^{-\delta}
+2iP_{\alpha\beta}\epsilon^{-\beta}\;.$$

2) Left superdiffeomorphisms. These may be written as
active\footnote{Their passive form acting on the coordinates of the
world-sheet superspace can be found in \cite{a2}. There a detailed
discussion of the $N=4$ world-sheet superconformal group is given
(see also \cite{a5}).} transformations on the fields as follows:
\begin{equation} \delta X^{\alpha\beta}=iD^+\bar D^+ (\Lambda^{(-)}
D^{--} X^{\alpha\beta})\;,\ \ \delta\Theta^{+\alpha}=iD^+\bar D^+
(\Lambda^{(-)} D^{--} \Theta^{+\alpha})\;, \label{diff}
\end{equation} where $\Lambda^{(-)}$ is a $u$-independent
($D^{++}\Lambda^{(-)} = 0$) superfield satisfying the constraint
\begin{equation} D^+\bar D^+\Lambda^{(-)}=0\;.  \label{cons}
\end{equation}
 Notice that the $\xi^{(+)}$-dependence of $\Lambda^{(-)}$ is not
restricted just as in the cases considered previously.  The
components of $\Lambda^{(-)}$ are easily shown to be the parameters
of left diffeomorphisms, local $N=4$ left supersymmetry and local
$SU(2)$ transformations (see \cite{a2}, \cite{a5}).  Assuming for the
time being that the Lagrange multipliers transform according to the
standard law \p{diff}, the variation of the action (\ref{action})
under these transformations is given, up to total harmonic
derivatives,  by \begin{equation} \delta S = i\int d^2\xi
[du]d^4\eta\; \partial_{(+)}\Lambda^{(-)} D^{--}X^{\alpha\beta}
\Theta^{+\gamma}\Theta^{+\delta}
\epsilon_{\alpha\beta\gamma\delta}\;.  \label{transf} \end{equation}
One has to work a little in order to show that this expression can be
compensated by choosing the full transformations of the Lagrange
multipliers to be as follows:  \begin{eqnarray} &&\delta
P_{\alpha\beta}=iD^+\bar D^+ \left[ \Lambda^{(-)} D^{--}
P_{\alpha\beta}-\epsilon_{\alpha\beta\gamma\delta}\;
\partial_{(+)}\Lambda^{(-)} D^{--}
(\Theta^{+\gamma}D^{--}\Theta^{+\delta})\right] \nonumber\\ && \delta
Q^-_\alpha =iD^+\bar D^+ [ \Lambda^{(-)} D^{--}
Q^-_\alpha+\epsilon_{\alpha\beta\gamma\delta}\;
\partial_{(+)}\Lambda^{(-)}((D^{--})^2X^{\beta\gamma}
\Theta^{+\delta} \\ &&
 -{2\over 3} \Theta^{+\beta}\Theta^{+\gamma}
(D^{--})^3\Theta^{+\delta} - {2\over 9}(D^{--})^3
(\Theta^{+\beta}\Theta^{+\gamma}\Theta^{+\delta})) ]\;.
\label{diffp}\nonumber \end{eqnarray}

 3) Right conformal invariance. The action is invariant under right
conformal transformations with parameter $\Lambda^{(+)}(\xi^{(+)})$
provided the fields $X$ and $\Theta$ transform as scalars and the
Lagrange multipliers transform as densities. Like in the cases $D=3$
and $D=4$, this invariance can be promoted to a right
superdiffeomorphism one, this time by changing
$\Lambda^{(+)}(\xi^{(+)})$ to a general analytic superfield, if one
introduces an analytic einbein $\chi^{++(+)}$ and replaces the
harmonic derivative $D^{++}$ by a covariant one:
$$D^{++}\rightarrow{\cal D}^{++}=D^{++}+\chi^{++(+)} \partial_{(+)}$$
with $\delta \chi^{++(+)}=-{\cal D}^{++}\Lambda^{(+)}+ \Lambda^{(+)}
\partial_{(+)}\chi^{++(+)}$. The action (\ref{action}) with this
replacement made is still globally space-time supersymmetric, and
invariant under left superdiffeomorphisms, provided the harmonic
derivative $D^{--}$ in (\ref{diff})  is replaced by a covariant one:
$$D^{--}\rightarrow{\cal D}^{--}=D^{--}+\chi^{--(+)} \partial_+$$
with:  $$\{{\cal D}^{++},{\cal D}^{--}\} =D^0 \Rightarrow {\cal
D}^{++}\chi^{--(+)}-{\cal D}^{--}\chi^{++(+)}=0.$$ This equation
determines $\chi^{--(+)}$ as a function of $\chi^{++(+)}$. The
parameter $\Lambda^{(-)}$ in (\ref{diff}), (\ref{diffp}) is now
covariantly $u$-independent:  $${\cal D}^{++}\Lambda^{(-)}={\cal
D}^{--}\Lambda^{(-)}=0.$$ In the Wess-Zumino gauge, the only
surviving component of $\chi^{++(+)}$ is $$\mu(x)=i\int [du]D^-\bar
D^-\chi^{++(+)} \vert_{\eta = 0} ,$$ which transforms under right
diffeomorphisms as $$\delta \mu(x)=-\partial_{(-)}\Lambda^{(+)}(x)-
\mu(x) \partial_{(+)}\Lambda^{(+)}(x)-\Lambda^{(+)}(x)\partial_{(+)}
\mu(x).$$

\subsection{Component action}

In order to find out the component content of the superfield action
\p{action} one has first to get rid of the harmonic dependence of the
superfields in it. This is achieved using the harmonic constraints
introduced in the action with the Lagrange multipliers $P$ and $Q$
(see \cite{a2} for the details).  Eliminating those infinite sets of
auxiliary fields and using the Wess-Zumino gauge for $\chi^{++(+)}$
just discussed, we are left with the following component fields:
$$x^{\alpha\beta}(\xi)=\int [du]X^{\alpha\beta}\vert_{\eta=0}\;,
\ \ \ \theta^{i\alpha} (\xi)=2\int [du]u^{-i}
\Theta^+\vert_{\eta=0}\;,$$ $$\lambda^\alpha =\int
[du]D^-\Theta^+\vert_{\eta=0}\;,\ \ \ \bar\lambda^\alpha =\int
[du]\bar D^-\Theta^+\vert_{\eta=0}\;,$$ $$\sigma_{\alpha\beta}=\int
[du]P_{\alpha\beta} \vert_{\eta=0}\;.$$
 Let us introduce a notation similar to the one used in section 2:
$$\pi^{\alpha\beta}_{(\pm)}= \partial_{(\pm)} x^{\alpha\beta}
+{i\over 2}\partial_{(\pm)}\theta^{i\alpha}\;\theta^{\beta}_i
-{i\over 2}\partial_{(\pm)}\theta^{i\beta}\;\theta^{\alpha}_i $$
$$\hat{\pi}^{\alpha\beta}_{(-)}= {\cal D}_{(-)} x^{\alpha\beta}
+{i\over2} {\cal D}_{(-)}\theta^{i\alpha}\;\theta^{\beta}_i
-{i\over2} {\cal D}_{(-)}\theta^{i\beta}\;\theta^{\alpha}_i \equiv
\pi^{\alpha\beta}_{(-)} + \mu\pi^{\alpha\beta}_{(+)}\;.$$ Here ${\cal
D}_{(-)} = -i \{[{\cal D}^{--}, D^+],\bar D^+\}$. Then the component
form of the action (\ref{action}) is given by:  \begin{eqnarray} S
&=& \int d^2\xi \left[\epsilon_{\alpha\beta\gamma\delta}
\;(2\pi^{\alpha\beta}_{(+)}\bar\lambda^\gamma\lambda^\delta
+2i\;\partial_{(+)}\theta^{i\alpha}\;\theta^{\beta}_i
\bar\lambda^\gamma \lambda^\delta \right. \nonumber \\
 &+& \left. i\;{\cal D}_{(-)} \theta^{i\alpha}\;
\theta^{\beta}_i\pi^{\gamma\delta}_{(+)})
-i\;\sigma_{\alpha\beta}\;(\hat{\pi}^{\alpha\beta}_{(-)}+2
\bar\lambda^\alpha\lambda^\beta)\right]\;.
\label{action6}\end{eqnarray} In components, the left local
supersymmetry transformations contained in \p{diff} read:
\begin{eqnarray} &&\delta\theta^{\alpha i}=\rho^i\bar\lambda^\alpha
+\bar\rho^i\lambda^\alpha \nonumber\\ &&\delta
x^{\alpha\beta}=-{i\over 2}\rho^i (\bar\lambda^\alpha\theta^\beta_i
-\bar\lambda^\beta\theta^\alpha_i) -{i\over 2}\bar\rho^i
(\lambda^\alpha\theta^\beta_i -\lambda^\beta\theta^\alpha_i)
\label{cotr}\\&& \delta\lambda^\alpha = i\rho^i\;{\cal
D}_{(-)}\theta^\alpha_i\;,\ \ \delta\bar\lambda^\alpha =
-i\bar\rho^i\;{\cal D}_{(-)}\theta^\alpha_i\nonumber\\&&
\delta\sigma_{\alpha\beta}= \epsilon_{\alpha\beta\gamma\delta}\;
\partial_{(+)}(\rho^i\bar\lambda^\gamma\theta^\delta_i
+\bar\rho^i\lambda^\gamma\theta^\delta_i)\;. \nonumber \end{eqnarray}
In the action (\ref{action6}) the field $\sigma_{\alpha\beta}$ is a
Lagrange multiplier for the constraint:  \be\label{const}
\bar\lambda^\alpha\lambda^\beta -\bar\lambda^\beta\lambda^\alpha
=-\hat{\pi}^{\alpha\beta}_{(-)}\;, \ee which has as a consequence the
Virasoro constraint \begin{equation}
(\hat{\pi}_{(-)})^2=\epsilon_{\alpha\beta\gamma\delta}\;
\pi^{\alpha\beta}_{(-)}\pi^{\gamma\delta}_{(-)} =0
\label{light}\end{equation} (the second Virasoro constraint can be
obtained in precisely the same manner as in the case $D=3$, see
section 2). Further, only the product $\bar\lambda\lambda$ appears in
the action, so we can  use the constraint \p{const} to rewrite the
action as follows:  \be S=\int d^2\xi
\epsilon_{\alpha\beta\gamma\delta}\left[
-\pi^{\alpha\beta}_{(+)}\hat{\pi}^{\gamma\delta}_{(-)}
-i\;\partial_{(+)}\theta^{i\alpha}\theta^{\beta}_i
\hat{\pi}^{\gamma\delta}_{(-)} +i\;{\cal D}_{(-)}\theta^{i\alpha}
\theta^{\beta}_i\pi^{\gamma\delta}_{(+)} \right]\; .  \ee This,
together with the constraint (\ref{light}), is just the action for
the GS superstring in six dimensions.

Once again, the procedure of elimination of the twistor variables
$\bar\lambda\;, \lambda$ breaks the left local supersymmetry, but
some memory of it is kept, which is just kappa symmetry.  It emerges
after the replacement
 $\rho^i =\lambda^\alpha \kappa^i_\alpha $, $\bar\rho^i
=-\bar\lambda^\alpha \kappa^i_\alpha$ in eqs. (\ref{cotr}) and the
elimination of the product $\bar\lambda\lambda$ from the resulting
transformations with the help of the on-shell constraint \p{const}.
In this way one obtains $$\delta\theta^{\alpha i} ={1\over
2}\kappa^i_\beta \pi_{(-)}^{\alpha\beta}\;,\ \ \delta
x^{\alpha\beta}=-{i\over 2}(\delta\theta^{\alpha i} \theta_i^\beta
-\delta\theta^{\beta i} \theta_i^\alpha )\;. $$ These can be
recognized as the kappa symmetry transformations of the GS
superstring. We stress again that this identification is only
possible on shell, where the constraint \p{const} is valid.

\section{Conclusions}

The obvious question which did not find its answer in the present
paper, is how to approach the most interesting case of the $D=10$
superstring with $N=(8,0)$ world-sheet supersymmetry. The problem is
that the notion of complex structure and the associated Grassmann
analyticity, heavily used in this paper, do not have a natural
extension to the case $D=10, N=(8,0)$. However, there exists an
alternative approach, which is based on the properties of the
eight-sphere realized as a coset space of the $D=10$ lorentz group,
and does not make use of any complex structures (see \cite{100} for
the case of the superparticle and a forthcoming publication for the
superstring).

Amongst possible immediate developments of the results presented here
let us mention coupling the above superstring actions to target-space
background supergravity and super-Yang-Mills, as well as introducing
additional world-sheet superfields in order to describe the internal
degrees of freedom of the heterotic superstrings.

Finally, let us point out that in the present paper we dealt with an
$n=1$ target superspace. In the context of a twistor-like formulation
this naturally goes together with an world-sheet supersymmetry of the
heterotic $(N,0)$ type. Analogous formulations for the $n=2$ GS
superstrings are expected to be combined with world-sheet
supersymmetries of the $(N,N)$ type and thus to involve two
independent sets of twistor variables \cite{a00}.  The latter may be
used to replace both vectors $\pi^{\mu}_{(+)}\;,$ $\pi^{\mu}_{(-)}$
and thus to simultaneously solve both Virasoro constraints of the
superstring, along the lines of refs.  \cite{{a00},{a0},{a13}}.

\end{document}